\documentclass[10pt,journal,twocolumn,twoside]{IEEEtran} % formal style

\usepackage{graphicx}
\usepackage{epstopdf}
\usepackage{float}
\usepackage{algorithmic}
\usepackage{array}
\usepackage{amsmath}
\usepackage{amssymb}
\usepackage{mdwmath}
\usepackage{mdwtab}
\usepackage{eqparbox}
\usepackage{stfloats}
\usepackage{fixltx2e}
\usepackage{cases}

\usepackage{xcolor}
\usepackage{makecell}

\usepackage[boxed,ruled,commentsnumbered]{algorithm2e}
\usepackage{url}
\usepackage{cite}
\ifCLASSOPTIONcompsoc
\usepackage[caption=false,font=normalsize,labelfont=sf,textfont=sf]{subfig}
\else
\usepackage[caption=false,font=footnotesize]{subfig}
\fi
\allowdisplaybreaks[4]

\newtheorem{theorem}{\bf Theorem}
\newtheorem{proposition}{Proposition}

\newtheorem{remark}{Remark}

\newcommand{\theoremref}[1]{Theorem~\ref{#1}}

\newcommand{\diag}{\mathop{\mathrm{diag}}}
\newcommand{\tr}{\mathop{\mathrm{trace}}}
\makeatletter
\renewcommand*{\@opargbegintheorem}[3]{\trivlist
      \item[\hskip \labelsep{\bfseries #1\ #2}] \textbf{(#3):}\ }
\makeatother

\begin{document}

\title
{Green Holographic MIMO Communications With A Few Transmit Radio Frequency Chains}
\author{ Shuaishuai Guo,~\IEEEmembership{Senior Member, IEEE}, Jia Ye,~\IEEEmembership{Member, IEEE}, \\Kaiqian Qu,~\IEEEmembership{Graduate Student Member, IEEE}, and Shuping Dang,~\IEEEmembership{Member, IEEE}
\thanks{
The work is supported in part by the National Natural Science Foundation of China under Grant 62171262 and Grant 62301090; in part by Shandong Provincial Natural Science Foundation under Grant ZR2021YQ47 and ZR2021LZHP003; in part by the Taishan Young Scholar under Grant tsqn201909043; in part by Major Scientific and Technological Innovation Project of Shandong Province under Grant 2020CXGC010109.
(\emph{Corresponding author: Jia Ye}).}
\thanks{ Shuaishuai Guo and Kaiqian Qu are  with the School of Control Science and Engineering, Shandong University, Jinan 250061, China, and also with Shandong Provincial Key Laboratory of Wireless Communication Technologies (e-mail: shuaishuai\textunderscore guo@sdu.edu.cn, qukaiqian@mail.sdu.edu.cn).}
\thanks{Jia Ye is with the School of Electrical Engineering, Chongqing University, Chongqing, 400044, China (yejiaft@163.com).}
\thanks{Shuping Dang is with the Department of Electrical and Electronic Engineering, University of Bristol, Bristol BS8 1UB, UK (e-mail: shuping.dang@bristol.ac.uk).}
   }
\maketitle

%\newpage

\begin{abstract}
Holographic multiple-input multiple-output (MIMO) communications are widely recognized as a promising candidate for the next-generation air interface.  With holographic MIMO surface, the number of the spatial degrees-of-freedom (DoFs) considerably increases and also significantly varies as the user moves. To fully employ the large and varying number of spatial DoFs, the number of equipped RF chains has to be larger than or equal to the largest number of spatial DoFs. 
However, this causes much waste as radio frequency (RF) chains (especially the transmit RF chains) are costly and power-hungry. To avoid the heavy burden, this paper investigates green holographic MIMO communications with a few transmit RF chains under an electromagnetic-based communication model. We not only look at the fundamental capacity limits but also propose an effective transmission, namely non-uniform holographic pattern modulation (NUHPM), to achieve the capacity limit in the high signal-to-noise (SNR) regime. 
The analytical result sheds light on the green evaluation of MIMO communications, which can be realized by increasing the size of the antenna aperture without increasing the number of transmit RF chains. Numerical results are provided to verify our analysis and to show the great performance gain by employing the additional spatial DoFs as modulation resources.

\end{abstract}


\section{Introduction}
\IEEEPARstart{M}{assive} multiple-input multiple-output (MIMO) as a mature technology has been applied in the fifth-generation (5G) New Radio (NR) standard \cite{BJORNSON2019}. With the application of massive MIMO and higher carrier frequency, a 5G base station consumes about three times the power of a fourth-generation (4G) base station but only covers around the one-third area. Consequently, the deployment of 5G networks leads to huge power consumption, and excessive electricity bills,  making operators less motivated to upgrade their communication networks from 4G to 5G \cite{Guo5G2022}. The huge power consumption becomes the main obstacle to green and sustainable communication networks. Although the recently emerged technologies such as reconfigurable intelligent surface \cite{9133134,9097454,9254161}, backscatter communications \cite{boyer2013invited,Zhao2023} are energy-friendly, they still cannot slow down the growth trend of power consumption significantly. Investigation into the power consumption of a base station has shown that the radio frequency (RF) chains (especially the power amplifiers in transmit RF chains) are the main cause of high power consumption. Therefore,  it is essential to investigate next-generation MIMO technology that utilizes a few RF chains for green and sustainable evaluation \cite{Guo2019,Guo2020,Guo2020a,Guo2022b}.

 A natural question to be answered first is: \emph{what might be next-generation MIMO technology beyond massive MIMO?} A straightforward evaluation way is continuously increasing the antenna numbers subject to a size constraint. Then, it comes to the concept of holographic MIMO \cite{Huang2020,Dardari2021}, which is referred to communications between antenna arrays with infinite numbers of antennas in a compact space, also known as large intelligent surfaces (LISs) \cite{Hu2018} and continuous-aperture (CAP) MIMO \cite{Zhang2022}. The benefits of this promising candidate can be partly observed from the ``holographic" term literally representing “describe everything” \cite{Dardari2021}. The study of the holographic MIMO is a rapidly growing area of research, with scientists exploring how the electromagnetic characteristics of wireless propagation channels help to approach the fundamental performance limit.

\subsection{Related Work and Motivation}
Recently, a substantial amount of literature has emerged on the holographic MIMO communications field from several perspectives, including communication and channel modeling, capacity and degree-of-freedom (DoF) analysis, and DoF explorations. 

\subsubsection{Communication and Channel Modeling}
A major challenge in analyzing and understanding the fundamental limits of holographic MIMO is the lack of mathematically tractable and numerically reproducible channel models. To retain some semblance to physical reality, electromagnetic-compliant models were developed in \cite{Wang2022,pizzo2020degrees,Pizzo2020,Pizzo2022,Li2022} by representing the channel response through Fourier plane-wave series expansion. Specifically, Pizzo {\itshape et al} made a valuable contribution to this area, who build the Fourier plane-wave spectral representation for the small-scale fading in far-field \cite{pizzo2020degrees,Pizzo2020}. The scattering environment, the number of radiative elements, and the structure of the receiving array are additionally considered in their extended work \cite{Pizzo2022} to comply with the physics of practical wave propagation better. Moreover, the electromagnetic-compliant channel model of the multi-user holographic MIMO communication systems was investigated in \cite{Li2022}. The authors in \cite{Yuan2023,Wang2022} pay more attention to the mutual coupling effect caused by the close deployment of antenna elements. In \cite{Yuan2023}, the correlation function between two antennas was formulated by the most commonly-used Clarke’s model. As a result of the mutual coupling, the distorted antenna patterns and antenna efficiency reduction were considered in \cite{Wang2022} to build up a more realistic channel model. 

In addition to the Fourier plane-wave spectral series-based electromagnetic-compliant channel models, an electromagnetic-based communication model between two spatially-continuous volumes introduced in \cite{miller_communicating_2000} is widely adopted in the existing literature \cite{Yuan2022,Torcolacci2022,Williams2020} to match the realistic continuous electromagnetic field. This kind of model is built upon Maxwell's equations and Green functions, which deliver the input-output relationship in the electric field. In fact, the Fourier spectral method is suitable for holographic MIMO systems in the far field, while the electromagnetic-based channel model fits more complex scenarios well \cite{Yuan2022}. However, this kind of model introduces integration operation in terms of the surface areas, thereby lacking mathematical tractability. Target at the near-field propagation scenario and planar surface, and approximation but in closed-form input-output channel expression was derived in \cite{wei2022tri}. These proposed communication and channel models lay the foundation for the following performance analysis and system design for holographic MIMO communications. 

\subsubsection{Analysis on Spatial DoFs and Capacity}
Based on the above-discussed electromagnetic model, the fundamental limits of holographic MIMO, such as spatial DoFs and capacity are gradually revealed. The spatial DoF is generally defined as the number of eigenvalues of the channel matrix, which is directly related to the achievable capacity. 

The importance of DoFs for achieving high data rates and reliable communication in holographic MIMO networks stimulates the literate growth in this field. For example, the research in \cite{pizzo2020degrees,yuan2021approaching, Torcolacci2022,sanguinetti_wavenumber-division_2022,Yuan2023} explored the channel spatial DoF in isotropic scattering environments, taking into account spatially-constrained apertures. The results demonstrated that the spatial DoF is proportional to the surface area, which differs from traditional analysis methods. Pizzo {\itshape et al} derived the conclusion that the 3D volumetric array does not offer any additional DoF compared to a 2D planar array \cite{Pizzo2022,pizzo2020degrees}. However, these works were conducted under the far-field propagation assumption, which is unrealistic especially given that the use of a holographic surface with numerous patch antennas is expected to facilitate near-field communication. The DoF evaluation in near-field communications is also of great importance, which attracts some researchers' attention. The authors in \cite{ji2023extra} showed that an extra 30\% DoFs can be achieved in the near-field compared with the traditional far-field scenario. The additional gain is closely related to the wavelength, and distance between the transmitter and receiver planes. In \cite{Dardari2020}, the fundamental limits of LIS-based communications have been investigated, demonstrating how multiple communication modes, namely the channel's DoF, can be obtained in line-of-sight (LoS) and near-field regime by suitably designing the amplitude and phase profiles of the transmitting and receiving LISs. Specifically, communication modes can be described as a set of parallel and orthogonal spatial channels, defined at the electromagnetic level, thus enabling spatial multiplexing capabilities in the system. In addition, the concept of effective DoF was investigated in \cite{Yuan2022} as a mathematical tool to represent the equivalent number of single-input single-output (SISO) channels in a MIMO system. The effective DoF is equal to the optimal number of sources or receivers required to approach the performance limit and is directly related to the slope of spectral efficiency. The experiment results showed that the effective DoF increases rapidly at the beginning and then reaches the maximum as the number of transmitting/receiving points increases, regardless of the antenna's polarization dimension. 

The capacity is also a critical performance metric in wireless communications, which lays the foundation for system design and optimization, efficient resource allocation, performance evaluation, and regulatory compliance. The capacity limit of holographic MIMO communication systems thus was investigated in several works from the electromagnetic perspective based on the constructed electromagnetic-based communication model. The authors in \cite{Yuan2022, Pizzo2022} derived the capacity limit expression in terms of the DoF. It was shown in \cite{Pizzo2022} that the spatial DoFs determine the slope of capacity in the large signal-to-noise ratio (SNR) regime when the channel state information is perfectly known at the transmitter and receiver. The capacity of the downlink holographic multi-user MIMO communication system adopting different linear detectors, including maximum-ratio transmission and zero-forcing precoding, was analyzed in \cite{Li2022}. It reveals the achievable capacity performance can be impaired by the reduced spacing distance among antennas due to the generated stronger signal correlation. Under the consideration of evanescent waves and available instantaneous channel state at the receiver, the authors in \cite{ji2023extra} show additional capacity can be brought in the near-field region compared to the far-field one. The aforementioned works have demonstrated the influence of channel state information, antenna spacing distance, and signal propagation distance on the achievable capacity. These findings thereby offer valuable insights for the design of holographic  MIMO communication systems.

\subsubsection{Techniques to Explore the Performance Limits}
 In fact, the holographic beamforming design is the key factor to determine communication performance. In order to approach the ultimate capacity of holographic MIMO systems, various beamforming techniques have been investigated in recent years. The beamforming technique proposed in \cite{Pizzo2022} is named the wavenumber-division multiplexing scheme, which is similar to the conventional frequency-division multiplexing to some extent. Specifically, the proposed scheme modulates transmitted symbols on the different wavenumbers of the generating orthogonal patterns by Fourier basis functions respectively. The orthogonality makes symbols become distinguishable, but are specialized for LoS communication scenarios. The beamforming design for the electromagnetic-based communication model is a challenging task since the model involves integration over the holographic surface. To address this issue, the authors in \cite{zhang_pattern-division_2022} utilized Fourier series expansion and its finite-element approximation to transform the design of continuous pattern functions into the design of their projection lengths on finite orthogonal bases, thereby proposing the pattern-division multiplexing scheme maximizing the system sum-rate. Moreover, the application of orbital angular momentum (OAM) to holographic MIMO systems using large intelligent surfaces was studied in \cite{Torcolacci2022}. It was shown that the generated basis functions exploiting the unique OAM property achieve suboptimal but high DoF wireless communication.  Deng {\itshape et al} proposes a kind of space multiple access technique for reconfigurable holographic surfaces that utilizes the holographic pattern division \cite{Deng2022}, which is capable of generating multiple desired directional beams towards different receivers, thereby enhancing the system sum rate massive connectivity. Some of the researchers also focus on treating the mutual coupling effect when performing transceiver design. The authors in \cite{Williams2020} proposed a novel matched filter-based transmitter design by utilizing the formulated expression of the radiated power that takes into account the mutual coupling matrix. This design employs the inverse of the mutual coupling matrix and provides improved directivity compared to the conventional design, with a directivity gain equaling the number of antennas.

In a short summary, current research on holographic MIMO focuses more on analyzing and digging into the greatly increased spatial DoFs. It has the assumption that the transmitter has sufficient RF chains compared to the number of spatial DoFs. As the DoFs vary as the transceiver moves, the equipped number of RF chains has to be larger than or equal to the largest number of spatial DoFs. {Present analyses indicate that in near-field scenarios, the spatial DoFs can reach up to the hundreds, implying the
deployment of numerous costly and power-intensive RF chains.} We believe it is not a sustainable evaluation. Thus, can we increase the antenna aperture without increasing the number of transmit RF chains to improve the spectral efficiency?
We refer to it as \emph{green holographic MIMO (GH-MIMO)} with a few transmit RF chains. In this work, we discuss its fundamental capacity limits and capacity-achieving transmission techniques  under an electromagnetic communication model.

\subsection{Contributions}
The contributions of this paper can be summarized as follows.
\begin{itemize}
    \item In this paper, we build the signal model for GH-MIMO with a few transmit RF chains and analyze its capacity under an electromagnetic model. The  capacity expression involves the entropy of Gaussian mixture distribution and high-dimensional integration makes it difficult to calculate.
    To improve the mathematical tractability, a closed-form  asymptotic capacity is derived.
    \item Based on the analyzed asymptotic capacity, we propose a capacity-achieving technique, namely non-uniform holographic pattern modulation (NUHPM). The optimal capacity-achieving holographic activation probability distribution and the optimal data stream input distribution are derived.
    In NUHPM, the additional spatial DoFs are employed as modulation resources. It can flexibly employ the varying number of DoFs when the transceiver moves with a fixed small number of RF chains. 
    \item  
     Both theoretical analysis and simulation results show the proposed NUHPM can provide a significant increase in spectral efficiency compared to only selecting the best holographic pattern for data transmission. Simulation results also show the effectiveness of increasing the aperture of antennas without increasing the number of transmit RF chains in capacity improvement.
\end{itemize}

\subsection{Organization}
The remainder of this paper is organized as follows. We describe the system model of a point-to-point GH-MIMO with a few transmit RF chains in Section II. Then, we show the number of spatial DoFs in GH-MIMO as well as their variation along with the change in communication distance. In Section III, we first introduce the conventional solution to exploit the spatial DoFs. Then we present a new understanding of using the holographic pattern as additional modulation resources and analyze its capacity limits and asymptotic capacity. In Section IV,  we propose a NUHPM to approach asymptotic capacity and compute the optimal holographic activation probability distribution and the optimal data stream input distribution. Section V provides numerical results that demonstrate the advantage of GH-MIMO with NUHPM. Conclusions are drawn in Section VI.

\subsection{Notations}
%%%%%%%%%%Notation%%%%%%%%%
Through the paper, we use $x$, $\mathbf{x}$, $\mathbf{X}$, $\mathcal{X}$ to denote a scaler, a vector, a matrix, and a set, respectively.  $||\mathbf{x}||_0$ and $||\mathbf{x}||$  denot the $\ell_0$ norm and $\ell_2$ norm of $\mathbf{x}$, respectively. $|\mathcal{X}|$ represents the size of set $\mathcal{X}$. $(\cdot)^T$, $(\cdot)^H$, $(\cdot)^*$, $(\cdot)^{\dag}$ and $\rm{det}(\cdot)$ stand for transpose, Hermitian transpose, complex conjugate, Moore-Penrose pseudo-inverse and  determination of the matrices, respectively. $\mathbb{C}$ denotes the set of complex numbers; $\int_{\mathcal{X}}d\mathbf{x}$ represents an integral operation on the set $\mathcal{X}$ with respect to the variable $\mathbf{x}$, and $\bigtriangledown_{\mathbf{x}}$ denoted the gradient operation with respect to vector $\mathbf{x}$;  $\mathsf{H}(\mathbf{x})$ stands for the entropy of $\mathbf{x}$; $\mathsf{I}(\mathbf{x};\mathbf{y})$ means the mutual information between $\mathbf{x}$ and $\mathbf{y}$; The trace, determinant, logarithm, and stochastic expectation are written as ${\rm{trace}}\left(\cdot\right)$, ${\rm{det}}\left(\cdot\right)$, ${\rm{log}}\left(\cdot\right)$ and $\mathbb{E}\{\cdot\}$. Finally, $\mathbf{I}_N$ denotes an $N\times N$ identity matrix, $f(\mathbf{x})$ stands for the probability density function (PDF) of $\mathbf{x}$.

\section{System Model}
\begin{figure*}[htpb]
 \centering
 \includegraphics[width=1\linewidth]{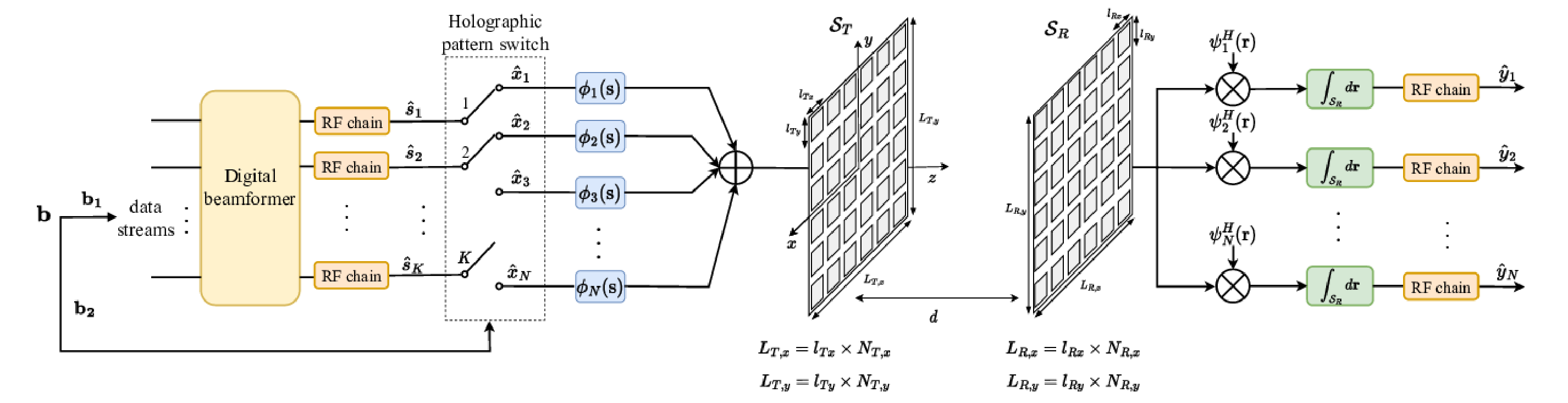}
 \caption{Schematic representation of a green holographic MIMO communications system spaced parallel of $d$ meters in the $z$-axis.}
 \label{System_Model}
\end{figure*}
As illustrated in Fig. \ref{System_Model}, we consider a GH-MIMO communication system with a few transmit RF chains, whose number is denoted by $K$. Both transmitter and receiver are equipped with holographic MIMO surfaces, denoted by $\mathcal{S}_T$ and $\mathcal{S}_R$, respectively. It is assumed the surfaces are rectangular of size $\mathcal{A}_T = L_{T,x}\times L_{T,y}$ and $\mathcal{A}_R = L_{R,x}\times L_{R,y}$, where $L_{P,x}$ and $L_{P,y}$, $P \in \{T,R\}$, denote surfaces’ horizontal and vertical lengths. There are $N_T = N_{T,x} \times N_{T,y}$ and $N_R = N_{R,x} \times N_{R,y}$ patch antennas on the transmitting and receiving holographic MIMO surfaces, where $N_{P,x}$ and $N_{p,y}$ represent the antenna number counted along horizontal and vertical directions, separately. Without loss of generality, we consider each patch antenna can transmit signals in three polarizations. The polarization dimension is generally reduced to two or even simplified to one in far-field communication scenarios, which is widely assumed in the existing literature. Accordingly, the radiated electric field $\mathbf{e}(\mathbf{r})$ at the location $\mathbf{r}\in\mathcal{S}_R$ resulting from the current $\mathbf{j}(\mathbf{s})$ generated at point $\mathbf{s}$ can be calculated by the dyadic Green’s function theorem as follows \cite{gruber2008new}
\begin{equation}
\mathbf{e}(\mathbf{r})=\int_{\mathcal{S}_T}\mathbf{G}(\mathbf{r},\mathbf{s})\mathbf{j}(\mathbf{s})d\mathbf{s},
\end{equation}
where $\mathbf{G}(\mathbf{r},\mathbf{s})$ is the dyadic Green’s function given by \cite{arnoldus2001representation}.
\begin{equation}
\begin{split}
\mathbf{G}(\mathbf{r},\mathbf{s})&=\frac{e^{j\kappa||\mathbf{r}-\mathbf{s}||}}{4\pi||\mathbf{r}-\mathbf{s}||}\left(\mathbf{I}+\frac{\bigtriangledown_{\mathbf{r}}\bigtriangledown_{\mathbf{r}}^H}{\kappa^2}\right) \\
&=\left.\frac{e^{j\kappa||\mathbf{r}-\mathbf{s}||}}{4\pi||\mathbf{r}-\mathbf{s}||}\right[\left(\mathbf{I}-\hat{\mathbf{p}}\hat{\mathbf{p}}^H\right)+\frac{j}{\kappa||\mathbf{r}-\mathbf{s}||}(\mathbf{I}
-3\hat{\mathbf{p}}\hat{\mathbf{p}}^H)\\
&\left.-\frac{1}{(\kappa||\mathbf{r}-\mathbf{s}||)^2}(\mathbf{I}-3\hat{\mathbf{p}}\hat{\mathbf{p}}^H)\right].
\end{split}
\end{equation}
and $\hat{\mathbf{p}}=\mathbf{p}/||\mathbf{p}||$ with $\mathbf{p}=\mathbf{r}-\mathbf{s}$.

Assuming the size of each transmitting patch antenna is $l_{T_x}\times l_{T_y}$, the radiated electric field $\mathbf{e}(\mathbf{r})$ thus can be recalculated as
\begin{equation}
\mathbf{e}(\mathbf{r})=\sum_{n_t = 1}^{N_T}\int_{-l_{T_x}/2}^{l_{T_x}/2}\int_{-l_{T_y}/2}^{l_{T_y}/2}\mathbf{G}(\mathbf{r},\mathbf{s}_{n_t}(x,y))\mathbf{j}(\mathbf{s}_{n_t}(x,y))dxdy. 
\end{equation}
Here, $\mathbf{s}_{n_t}(x,y) = [s_x-x,s_y-y,s_z]^T$ related to the known transmitting position value at the $n_t$-th patch antenna, that is, $\mathbf{s}_{n_t}= [s_x,s_y,s_z]^T$. Under the assumption that the current distribution $\mathbf{j}(\mathbf{s}_{n_t}(x,y))$ is constant. The radiated electric field $\mathbf{e}(\mathbf{r})$ can be obtained in approximated closed form or calculated numerically. Similarly, assuming the size of each receiving patch antenna is $l_{R_x}\times l_{R_y}$, it is widely assumed that the power received by each receiving patch antenna will be proportional to its receiving area as the transmitting holographic MIMO surface is much bigger than each receiving antenna patch \cite{wei2022tri}. Therefore, the 
channel between each $n_r$-th receive and each $n_t$-th transmit patch antennas can be expressed as
 \begin{equation}
\mathbf{H}_{n_r,n_t}= l_{R_x}l_{R_y} \int_{-l_{T_y}/2}^{l_{T_y}/2}\int_{-l_{T_x}/2}^{l_{T_x}/2}\mathbf{G}(\mathbf{r}_{n_r},s_{n_t}(x,y))dxdy,
\end{equation}
which is of $3\times 3$ dimension matrix. This leaves us with the input-output relationship of the holographic MIMO communication system as
\begin{equation}
\mathbf{y}_{n_r}=\sum_{n_t=1}^{N_T}\mathbf{H}_{n_r,n_t}\mathbf{x}_{n_t}+\mathbf{z}_{n_r},
\end{equation}
where $\mathbf{x}_{n_t} \in \mathbb{C}^{3}$ is the transmitted signal vector, $\mathbf{y}_{n_r}\in \mathbb{C}^{3}$ is the received signal vector, and $\mathbf{z}_{n_r}\in \mathbb{C}^{3}$ denotes the additive Gaussian noise with zero mean and covariance $\sigma^2\mathbf{I}_{3}$.

Stacking all $\mathbf{x}_{n_t}$, $\mathbf{y}_{n_r}$,   $\mathbf{z}_{n_r}$ into vectors $\mathbf{x}=[\mathbf{x}_1^T,\mathbf{x}_2^T,\cdots,\mathbf{x}_{N_T}^T]^T\in\mathbb{C}^{3N_T\times 1}$, $\mathbf{y}=[\mathbf{y}_1^T,\mathbf{y}_2^T,\dots,\mathbf{y}_{N_R}^T]^T\in\mathbb{C}^{3N_R\times 1}$, and $\mathbf{z}=[\mathbf{z}_1^T,\mathbf{z}_2^T,\cdots,\mathbf{z}_{N_R}^T]^T\in\mathbb{C}^{3N_R\times 1}$ generates
\begin{equation}
\mathbf{y}=\mathbf{H}\mathbf{x}+\mathbf{z},
\end{equation}
where $\mathbf{H}\in\mathbb{C}^{3N_R\times 3N_{T}}$ represents the equivalent channel, with $n_r$th row $n_t$th column sub-matrix being $\mathbf{H}_{n_r,n_t}$. 

Let $N$ represent the rank of $\mathbf{H}$. To separate the cross-talk channels, we introduce singular vector decomposition (SVD) and use the orthogonal basis as the transmitting and receiving holographic patterns. As shown in Fig. \ref{System_Model}, we use $\phi_{1}(\mathbf{s}), \phi_2(\mathbf{s}),\cdots,\phi_{N}(\mathbf{s})\in\mathbb{C}^{3N_T\times 1}$ to denote the orthogonal transmitting holographic patterns and $\psi_{1}(\mathbf{s}), \psi_2(\mathbf{s}),\cdots,\psi_{N}(\mathbf{s})\in\mathbb{C}^{3N_R\times 1}$ to denote the orthogonal receiving holographic patterns, and then we can rewrite the channel as
\begin{equation}
\mathbf{H}=\pmb{\Psi}\pmb{\Lambda}\pmb{\Phi}^H
\end{equation}
where $\pmb{\Phi}=[\phi_{1}(\mathbf{s}), \phi_2(\mathbf{s}),\cdots,\phi_{N}(\mathbf{s})]\in\mathbb{C}^{3N_T\times N}$, $\pmb{\Psi}=[\psi_{1}(\mathbf{s}), \psi_2(\mathbf{s}),\cdots,\psi_{N}(\mathbf{s})]\in\mathbb{C}^{3N_R\times N}$, and $\pmb{\Lambda}\in\mathbb{C}^{N\times N}$ denote the diagonal singular-value matrix.

Utilizing $\hat{\mathbf{x}}\in\mathbb{C}^{N\times 1}$ and $\hat{\mathbf{y}}\in \mathbb{C}^{N\times 1}$ to represent the input and output to the equivalent channel, the input-output relationship can be written as
\begin{equation}\label{GHMIMO}
\hat{\mathbf{y}}=\pmb{\Lambda}\hat{\mathbf{x}}+\hat{\mathbf{z}},
\end{equation}
where $\hat{\mathbf{y}}=\pmb{\Psi}^H\mathbf{y}$, $\mathbf{x}=\pmb{\Phi}\hat{\mathbf{x}}$,  $\hat{\mathbf{z}}=\pmb{\Psi}^H\mathbf{z}$, and $\mathbf{z}\sim\mathcal{CN}(\mathbf{0},\sigma_2\mathbf{I}_{N})$.

\begin{figure}
    \centering
    \includegraphics[width=1\linewidth]{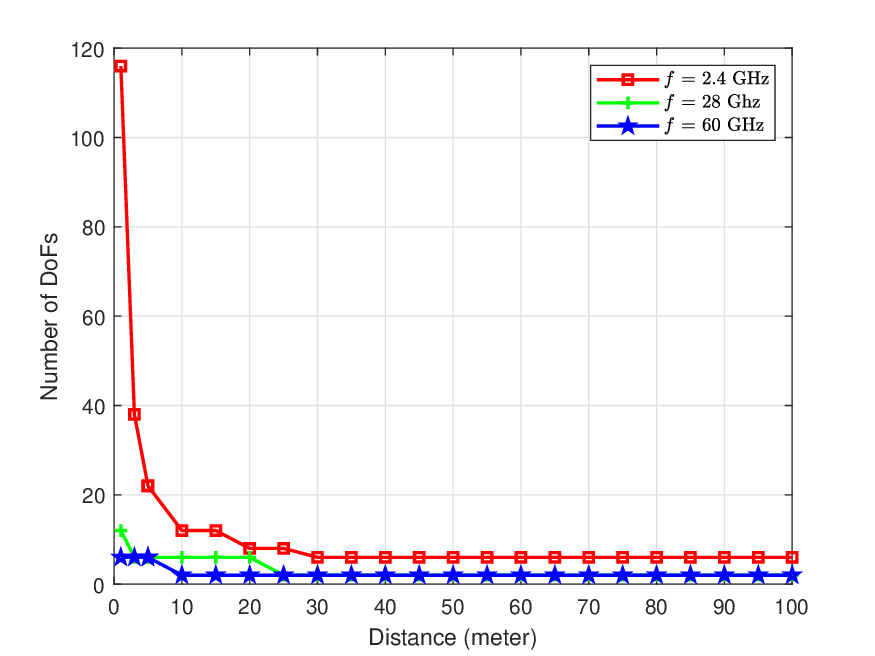}
    \caption{Number of DoFs of triple-polarized electromagnetic GH-MIMO models versus distance.}
    \label{dof}
\end{figure}

As we mentioned before, the spatial DoF is one of the most critical metrics to reveal the performance limits of electromagnetic GH-MIMO models. In order to demonstrate the properties of the considered system, numerical results with respect to the number of DoFs are illustrated in Fig. \ref{dof}. We consider a transceiver antenna surface placed parallel to each other configured with $l_{T,x}=l_{T,y}=l_{R,x}=l_{R,y}=0.4\lambda$, $N_{T,x}=N_{T,y}=N_{R,x}=N_{R,y}=16$. Also, We assume the antenna array is in the x-y plane, which is perpendicular to  the z-axis. The number of DoFs of triple-polarized electromagnetic GH-MIMO models varies with distance, as shown in Fig. \ref{dof}. Moreover, We
illustrate the variation of DoFs at three frequencies of $2.4$ gigahertz (GHz), $28$ GHz, and $60$ GHz. It can be seen that the DoF increases sharply as the distance decreases, reaching up to $120$ when $f=2.4$ GHz.
\remark{For the considered triple-polarized electromagnetic GH-MIMO models, one can observe that the number of spatial DoFs is huge and also varies as the communication distance varies. To employ the large and varying number of spatial DoFs, it is best if the transceivers are equipped with enough RF chains. That is, the number of RF chains is greater than the largest number of spatial DoFs. This is a costly solution. In this paper, we focus more on the realistic scenario with only a few transmit RF chains.}

% From  triple-polarized electromagnetic GH-MIMO models, one can observe that the number of spatial DoFs is huge and also varies as the communication distance varies. To demonstrate the phenomenon, numerical results are illustrated in Fig. \ref{dof}. We consider a transceiver antenna surface placed parallel to each other configured with $l_{T,x}=l_{T,y}=l_{R,x}=l_{R,y}=0.4\lambda$, $N_{T,x}=N_{T,y}=N_{R,x}=N_{R,y}=16$. Also, We assume the antenna array is in the x-y plane, which is perpendicular to  the z-axis. The number of DoFs of triple-polarized electromagnetic GH-MIMO models varies with distance, as shown in Fig. \ref{dof}. Moreover, We
% illustrate the variation of DoFs at three frequencies of $2.4$ gigahertz (GHz), $28$ GHz, and $60$ GHz. It can be seen that the DoF increases sharply as the distance decreases, reaching up to $120$ when $f=2.4$ GHz.
% To employ the large and varying number of spatial DoFs, it is best if the transceivers are equipped with enough RF chains. That is, the number of RF chains is greater than the largest number of spatial DoFs. This is a costly solution. In this paper, we focus more on the realistic scenario with only a few transmit RF chains.

\section{Capacity Limits and  DoFs Exploitation With A Few Transmit RF Chains}
If the number of transmit RF chains $K$ is much less than the number of spatial DoFs, the number of data streams is limited. This is equivalent to imposing a sparsity constraint on the input of the GH-MIMO communication model, i.e.,
\begin{equation}
||\hat{\mathbf{x}}||_0=K.
\end{equation}
The $K$ non-zero elements in $\hat{\mathbf{x}}$ are the data symbols sent by transmit RF chains. 
The capacity of GH-MIMO becomes the maximum mutual information with any input distribution under not only a power constraint but also a sparsity constraint, which
mathematically described as
\begin{equation}\label{eq10}
C=\max_{{f(\hat{\mathbf{x}})\atop \mathbb{E}(\tr\{\hat{\mathbf{x}}\hat{\mathbf{x}}^H)\}\leq 1}\atop ||\hat{\mathbf{x}}||_0=K}\mathcal{I}
(\hat{\mathbf{x}},\hat{\mathbf{y}}|\pmb{\Lambda}).
\end{equation}
To solve this problem, we first transform the sparsity constraint into
\begin{equation}
\hat{\mathbf{x}}=\mathbf{E}_i\hat{\mathbf{s}_i},
\end{equation}
where $\mathbf{E}_i\in\mathbb{C}^{N\times K}$ denotes the selection matrix being composed of $K$ vector basis of $N$ dimension. $\hat{\mathbf{s}_i}\in\mathbb{C}^{K\times 1}$ represents the  data symbols sent by the transmit RF chains and obeys a complex Gaussian distribution with zero mean and covariance $\mathbf{Q}_i$ when $\mathbf{E}_i$ is activated, i.e., $\hat{\mathbf{s}_i}\sim \mathcal{CN}(\mathbf{0},\mathbf{Q}_i)$.  $\mathbf{E}_i$ determines the positions of the non-zero elements and plays the role of holographic pattern selection. As there are a total of $N$ vector basis of $N$ dimension, therefore the number of feasible candidates for $\mathbf{E}_i$ is $\left(N\atop K\right)$ and we denote the candidates set as $\mathcal{E}$. Based on this denotations, we have $|\mathcal{E}|=\left(N\atop K\right)$.

\subsection{Conventional Solution and Its Spectral Efficiency}
{The most popular solution to deal with insufficient RF chains is to select the best $K$ out of $N$ holographic patterns for MIMO communications, which is also widely recognized as the ``optimal" solution\cite{Guo2020,Guo2019}}. Here, we refer to it as the best holographic pattern selection (BHPS).

Mathematically, the selection can be conducted by
\begin{equation}\label{eqF}
\mathbf{E}_i^{\mathrm{opt}}=\arg\max_{\mathbf{E}_i\in\mathcal{E}}\log_2\det\left(\mathbf{I}_{N}+\frac{1}{\sigma^2}\pmb{\Lambda}\mathbf{E}_i\mathbf{Q}_i\mathbf{E}_i^H\pmb{\Lambda}^H\right).
\end{equation}
 The optimal solution to the problem (\ref{eqF}) is to choose 
the holographic patterns that 
correspond to the largest $K$ singular values. The optimal $\mathbf{Q}_i$ is the diagonal water-filling power allocation matrix. It is obvious that only the best $K$ spatial DoFs are employed and the other $(N-K)$ spatial DoFs are left unexplored.

\subsection{New Understanding and Its Capacity Limit}

By rethinking the concept of capacity in (\ref{eq10}),  we realize that the old understanding overlooks the information-carrying capacity of holographic pattern hopping. In fact, the holographic pattern activation uncertainty can also be employed to carry additional information.  Let the holographic pattern selection matrix $\mathbf{E}_i$ be activated with probability $p_i$. Considering  $\hat{\mathbf{y}}=\pmb{\Lambda}\mathbf{E}_i\hat{\mathbf{s}}+\hat{\mathbf{z}}$ and both $\hat{\mathbf{s}}$ and $\hat{\mathbf{z}}$ obeys complex Gaussian distribution, $\hat{\mathbf{y}}$ will obey a  Gaussian mixture distribution whose PDF can be expressed as
\begin{equation}
f(\hat{\mathbf{y}})=\sum_{i=1}^{|\mathcal{E}|}p_i f_i(\hat{\mathbf{y}}),
\end{equation}
and
\begin{equation}
f_i(\hat{\mathbf{y}})=\frac{1}{\pi^{N}\det(\mathbf{D}_i)}\exp\left(-\hat{\mathbf{y}}^H\mathbf{D}_i^{-1}\hat{\mathbf{y}}\right),
\end{equation}
with
\begin{equation}\label{eq15}
\mathbf{D}_i=\mathbf{I}_{N}+\frac{1}{\sigma^2}{\mathbf{\Lambda}}\mathbf{E}_i\mathbf{Q}_i\mathbf{E}_i^H{\mathbf{\Lambda}^H}.
\end{equation}
The mutual information can be expressed as
\begin{equation}\label{eq16}
\mathsf{I}(\hat{\mathbf{x}};\hat{\mathbf{y}}|\mathbf{\Lambda})=\mathsf{H}(\hat{\mathbf{y}})-\mathsf{H}(\hat{\mathbf{y}}|\mathbf{\Lambda},\hat{\mathbf{x}}).
\end{equation}
The first term of (\ref{eq16}) is the entropy of the Gaussian mixture, which can be expressed as
\begin{equation}
\begin{split}
\mathsf{H}(\hat{\mathbf{y}})&=\mathbb{E}[-\log_2 f({\hat{\mathbf{y}}})]\\
&=-\int_{\mathbb{C}^{K}}\log_2 f(\hat{\mathbf{y}})\sum_{i=1}^{|\mathcal{E}|}p_i f_i(\hat{\mathbf{y}}) d\hat{\mathbf{y}}.
\end{split}
\end{equation}
The second term of (\ref{eq16}) can be expressed
by
\begin{equation}
\begin{split}
\mathsf{H}(\hat{\mathbf{y}}|\mathbf{\Lambda},\hat{\mathbf{x}})&=\mathsf{H}(\hat{\mathbf{z}})=N\log_2(\pi e).
\end{split}
\end{equation}
Summarizing above, we have the expression of mutual information to be 
\begin{equation}
\mathsf{I}(\hat{\mathbf{x}};\hat{\mathbf{y}}|\mathbf{\Lambda})=-\int_{\mathbb{C}^{K}}\log_2 f(\hat{\mathbf{y}})\sum_{i=1}^{|\mathcal{E}|}p_i f_i(\hat{\mathbf{y}}) d\hat{\mathbf{y}}-K\log_2(\pi e).
\end{equation}
Then, the capacity limit can be expressed by
\begin{equation}\label{eq20}
\begin{split}
C=&\max_{\{p_i\},\{\mathbf{Q}_i\}}  \mathsf{I}(\hat{\mathbf{x}};\hat{\mathbf{y}}|\mathbf{\Lambda})\\
=&\max_{\{p_i\},\{\mathbf{Q}_i\}} -\int_{\mathbb{C}^{K}}\log_2 f(\hat{\mathbf{y}})\sum_{i=1}^{|\mathcal{E}|}p_i f_i(\hat{\mathbf{y}}) d\hat{\mathbf{y}}-N\log_2(\pi e),\\
\mathrm{s.~t.:}& \sum_{i=1}^{|\mathcal{E}|}p_i\tr(\mathbf{Q}_i)=1,~\sum_{i=1}^{|\mathcal{E}|}p_i=1.
\end{split}
\end{equation}
where the first constraint is the average power constraint since
\begin{equation}
\begin{split}
\tr\left(\hat{\mathbf{x}}\hat{\mathbf{x}}^H\right)&=\sum_{i=1}^{N_c}p_i\tr(\mathbf{E}_i\hat{\mathbf{s}}_i\hat{\mathbf{s}}_i^H\mathbf{E}_i^H)\\
&=\sum_{i=1}^{N_c}p_i\tr(\mathbf{Q}_i)\leq 1.
\end{split}
\end{equation}

\subsection{Asymptotic Capacity}
As we can observe from (\ref{eq20}), the mutual information involves the computation of the entropy of the Gaussian mixture distribution and the high-dimension integration make it difficult to calculate. To improve the mathematical tractability of the mutual information, we resort to analyzing the asymptotic property in high SNR regimes. As is known, the entropy of Gaussian mixture distribution is lower bounded by \cite{Huber2008,Ibrahim2016}
\begin{equation}\label{eq21}
\begin{split}
\mathsf{H}(\hat{\mathbf{y}})&\geq  -\sum_{i=1}^{|\mathcal{E}|}p_i\log_2\left(\sum_{j=1}^{|\mathcal{E}| }\frac{p_j}{\pi^{N}\det(\mathbf{D}_i+\mathbf{D}_j)}\right)\\
&\triangleq {\mathsf{H}_{LB}(\hat{\mathbf{y}})}.
\end{split}
\end{equation}
 Based on the analysis in \cite{He2018}, it has been proved that there is a constant gap between the lower bound and the real entropy in the high SNR regime, which is  $N(\log_2 e-1)$. It means
 \begin{equation}\label{eq22}
 \lim_{1/\sigma^2\rightarrow +\infty} \mathsf{H}(\hat{\mathbf{y}})= {\mathsf{H}_{LB}(\hat{\mathbf{y}})}+N(\log_2 e-1)
 \end{equation}
Based on the relationship, we have the following proposition.

 \begin{proposition}
 In the high SNR regime, the spectral efficiency of GH-MIMO using non-uniform holographic pattern modulation can be asymptotically written as
 \begin{equation}
 \begin{split}
\lim_{1/\sigma^2\rightarrow +\infty} \mathsf{I}(\hat{\mathbf{x}};\hat{\mathbf{y}}|\mathbf{\Lambda})=&\sum_{i=1}^{|\mathcal{E}|}p_i\log_2\det\left(\mathbf{D}_i\right)-\sum_{i=1}^{|\mathcal{E}|}p_i\log_2p_i.
\end{split}
 \end{equation}
 \end{proposition}
 \begin{IEEEproof}
 See Appendix A.
 \end{IEEEproof}

 Based on this proposition, we can obtain the asymptotic capacity by solving
 \begin{equation}\label{eq24}
 \begin{split}
 C_{\mathrm{asym}}=&\max_{\{p_i\},\{\mathbf{Q}_i\}} \sum_{i=1}^{|\mathcal{E}|}p_i\log_2\det\left(\mathbf{D}_i\right)-\sum_{i=1}^{|\mathcal{E}|}p_i\log_2p_i,
 \\
\mathrm{s.~t.:}& \sum_{i=1}^{|\mathcal{E}|}p_i\tr(\mathbf{Q}_i)=1,~\sum_{i=1}^{|\mathcal{E}|}p_i=1.
\end{split}
 \end{equation}

The key to obtaining the closed-form asymptotic capacity is to find the holographic pattern activation probability distribution $\{p_i\}$ and the covariance matrices of the channel input distributions $\{\mathbf{Q}_i\}$. In the following section, we will discuss the solutions.

\section{Capacity-Achieving Holographic Pattern Activation Probability Distribution and Data Stream Input Distribution}

The optimization problem (\ref{eq24}) can be solved by the Lagrange methods. The optimal solutions are concluded in the following theorem.
\begin{theorem}
 The optimal $\{\mathbf{Q}_i^*\}$ can be expressed  as
\begin{equation}
\mathbf{Q}_i^*=\diag({\sigma_{i1}^*},\cdots,{\sigma_{iK}^*}),
\end{equation}
where
\begin{equation}
\sigma_{i,j}^*=\left(\frac{1}{\xi_i\ln2}-\frac{\sigma^2}{\lambda_{ij}^2}\right)^+,~j=1,\cdots,K,
\end{equation}
in which  $\lambda_{ij}^2$ is the channel gain of the $j$-data stream when $\mathbf{E}_i$ is activated and $\xi$  satisfies
\begin{equation}
\sum_{j=1}^{K}\left(\frac{ 1}{\xi_{i}\ln2}-\frac{\sigma^2}{\lambda_{ij}^2}\right)^+=1.
\end{equation}
 The optimal $\{p_i^*\}$ can be given by
\begin{equation}\label{eq29}
p_i^*=\frac{\det(\mathbf{D}_i^*)}{\sum_{i=1}^{|\mathcal{E}|}\det(\mathbf{D}_i^*)},~i=1,\cdots,|\mathcal{E}|,
\end{equation}
where $\mathbf{D}_i^*=\mathbf{I}_{N}+\frac{1}{\sigma^2}{\mathbf{\Lambda}}\mathbf{E}_i\mathbf{Q}_i^*\mathbf{E}_i^H{\mathbf{\Lambda}^H}$.
\end{theorem}
\begin{IEEEproof}
See Appendix B.
\end{IEEEproof}
 {From (\ref{eq29}), it is observed that holographic pattern activation probabilities are non-equal. They are related to the equivalent channel conditions,  and better channels will be activated with higher probabilities.
To transform independent and Bernoulli($1/2$) distributed input bits into a sequence of output symbols with a desired distribution,  distribution matching techniques \cite{Schulte2016} can be adopted.} 
 It is totally different from the conventional solution that only activates the best holographic patterns.
\begin{theorem}\label{theo2}
With the optimal $\{p_i^*\}$ and $\{\mathbf{D}_i^*\}$, the capacity in the high SNR regime can be approximately expressed as 
\begin{equation}\label{CH}
 C_{\mathrm{asym}}=\log_2 \sum_{i=1}^{|\mathcal{E}|}\det{(\mathbf{D}_i^*)}.
\end{equation}
\end{theorem}
\begin{IEEEproof}
Substituting the optimal $\{p_i^*\}$ and $\{\mathbf{D}_i^*\}$  into (\ref{eq24}) yields
\begin{equation}
\begin{split}
C_{\mathrm{asym}}=&\sum_{i=1}^{|\mathcal{E}|}\frac{\det(\mathbf{D}_i^*)}{\sum_{i=1}^{|\mathcal{E}|}\det(\mathbf{D}_i^*)}\log_2\det\left(\mathbf{D}_i^*\right)\\&-\sum_{i=1}^{|\mathcal{E}|}\frac{\det(\mathbf{D}_i^*)}{\sum_{i=1}^{|\mathcal{E}|}\det(\mathbf{D}_i^*)}\log_2\frac{\det(\mathbf{D}_i^*)}{\sum_{i=1}^{|\mathcal{E}|}\det(\mathbf{D}_i^*)}\\
=&\log_2 \sum_{i=1}^{|\mathcal{E}|}\det{(\mathbf{D}_i^*)}.
\end{split}
\end{equation}
\end{IEEEproof}
Theorem 2 theoretically proves that the proposed NUHPM always outperforms BHPS, because BHPS has an apparently lower spectral efficiency of 
\begin{equation}\label{CB}
C^{\rm{BHPS}}=\log_2 \max_{i=1,2,\cdots,|\mathcal{E}|} \det{(\mathbf{D}_i^*)}.
\end{equation}

\remark{In NUHPM, one of the available holographic patterns will be selected from this candidate set based on its associated activation probability. To provide a clearer understanding of how this process carries additional information, we can assign distinct indices to each holographic pattern within the candidate set. At the receiver, these indices can be decoded, effectively revealing the selected pattern for each symbol duration. This decoded pattern activation information itself constitutes the additional information delivered by the system. In essence, this mechanism resembles beam hopping techniques, but with a unique twist: the pattern selection dynamically changes based on the additional information bits that need to be conveyed. Therefore, the proposed scheme leverages holographic pattern activation uncertainty to transmit not only the primary data payload but also the dynamically changing pattern activation information, effectively enhancing the spectral efficiency and information-carrying capacity of the system.  

There are two major challenges in implementing NUHPM. The first challenge is the acquirement of perfect channel state information. It requires a lot of communication resources and the channel estimation process is affected by channel noise, making the perfect channel estimation infeasible. In practical applications, NUHPM can be implemented under imperfect channel estimation, but with performance degradation. The performance achieved under perfect channel estimation in this paper will be the upper bound of practical implementations. The second challenge is that the fast beam hopping for carrying additional information requires fast RF switches. Conventional beam hopping is operated per coherent time slot, which is tens of milliseconds. In NUHPM, the beam hopping is operated per symbol time.  Soujeri and Kaddoum \cite{Ebrahim2016} have reviewed the switching speed of micro-electronic systems (MEMS), which is known to be in the range of 2-50 $\mu s$. This renders it inefficient for today's high-speed applications. With the rapid development of RF switches, parasitic antenna arrays for MIMO applications using semiconductor diodes faster than $0.1$ $\mu s$ (100 $ns$) are now available, making implementing NUHPM feasible for high-speed applications.
%The third challenge is using hybrid beamformers to approach the optimal beamformers in real implementation. This can be conducted by solving 
%\begin{equation}
%\begin{split}
%&\min_{\mathbf{F}_{RF}^i,\mathbf{F}_{BB}^i}||\pmb{\Phi}\mathbf{E}_i-\mathbf{F}_{RF}^i\mathbf{F}_{BB}^i||_{F}^2,\\
%&\mathrm{subject to}:~\mathbf{F}_{RF}^i\in \mathcal{U}^{N_t\times K},~ \mathbf{F}_{BB}\in\mathbf{}
%\end{split}
%\end{equation}

}

\section{Simulations and Discussions}
To validate our analysis and show the advantage of the proposed NUHPM,  we evaluate the spectral efficiency of the proposed NUHPM and the conventional BHPS as well as uniform holographic pattern modulation (UHPM) in this section. It is noteworthy that we use a very small number of RF links as a constraint. The simulation parameters for surface structure are the same as the ones set in Fig. \ref{dof}, the position of the first patch transmit antenna is set at the center of the coordinate, that is $[0,0,0]^T$, and the perpendicular distance between the transmitting surface and receiving surface is set as $d= 1$ m. The simulated signal propagation happens on the carrier frequency $f = 2.4$ GHz with a single RF chain unless otherwise specified. 

\subsection{Comparison With BHPS and UHPM}
\begin{figure}
    \centering
    \includegraphics[width=1\linewidth]{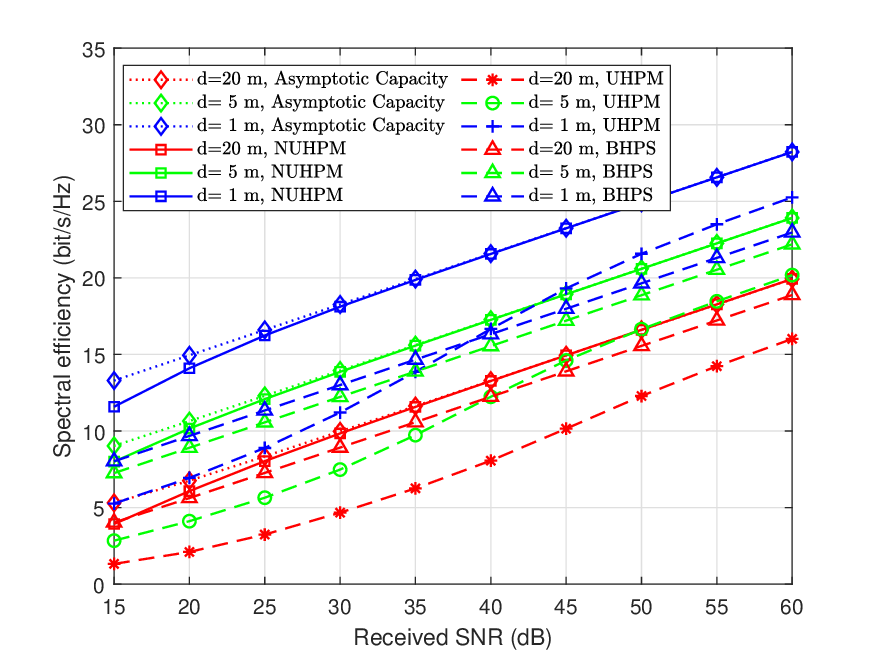}
    \caption{Spectral efficiency comparison among NUHPM, UHPM, and BHPS.}
    \label{snr}
\end{figure}

In the first experiment, we compare the proposed NUHPM scheme with the BHPS scheme and {the UHPM scheme} over different communication distances when the number of the transmitted RF chain $K=1$. As can be seen from Fig. \ref{snr}, NUHPM scheme always outperforms the BHPS scheme {and the UHPMs scheme}, which is in accordance with the theoretical analysis results. On the other hand, the cut of communication distance leads to higher achievable spectral efficiency. However, the improvement brought by the commonly adopted BHPS scheme is much lower than the one brought by the proposed NUHPM scheme, especially in the near-field region. Specifically, the BHPS scheme only provides around $1$ dB gain when $d$ reduces from $5$ m to $1$ m, while the NUHPM scheme offers about $12$ dB gains. On the other hand, the NUHPM scheme always outperforms the BHPS scheme. This finding is in accordance with the theoretical analysis results.  Moreover, the exact spectral efficiency\footnote{The exact spectral efficiency is calculated by Monte Carlo simulations.} represented by the line ’NUHPM’ coincides with the asymptotic spectral efficiency, especially in the high SNR regime, which verifies the correctness of \theoremref{theo2}.%our derivation.

\subsection{Impact of Communication Distance}
\begin{figure}
    \centering
    \includegraphics[width=1\linewidth]{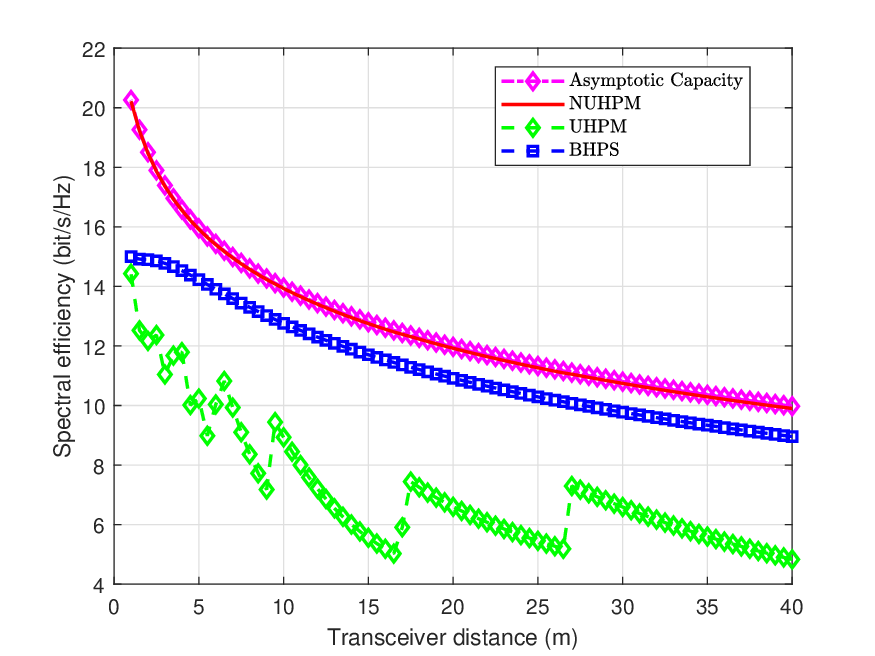}
    \caption{Spectral efficiency versus transceiver distances at an SNR of $20$ dB.}
    \label{distance}
\end{figure}

To better investigate the impact of the perpendicular distance between the transmitting surface and receiving surface on the achieved spectral efficiency, we simulate the spectral efficiency as a function of the communication distance $d$ at a high SNR of $20$ dB. The simulation results given in Fig. \ref{distance} that higher spectral efficiency can be achieved as the distance decreases. The NUHPM scheme and BHPS scheme bring 100\% and 57\% performance gains when the transceiver distance changes from $20$ m to $1$ m, respectively. Both of them greatly outperform the UHPM. It is noteworthy that the performance gain of NUHPM due to distance reduction becomes larger when the communication distance is less than around 5 meters, and is much higher than the one introduced by BHPS. These findings agree with the results observed in Fig. \ref{snr}. However, the performance variation tendency of NUHPM remains almost the same as the distance changes satisfying $r>5$ m, while BHPS owns these properties over the whole distance range. 
\begin{figure}
    \centering
    \includegraphics[width=1\linewidth]{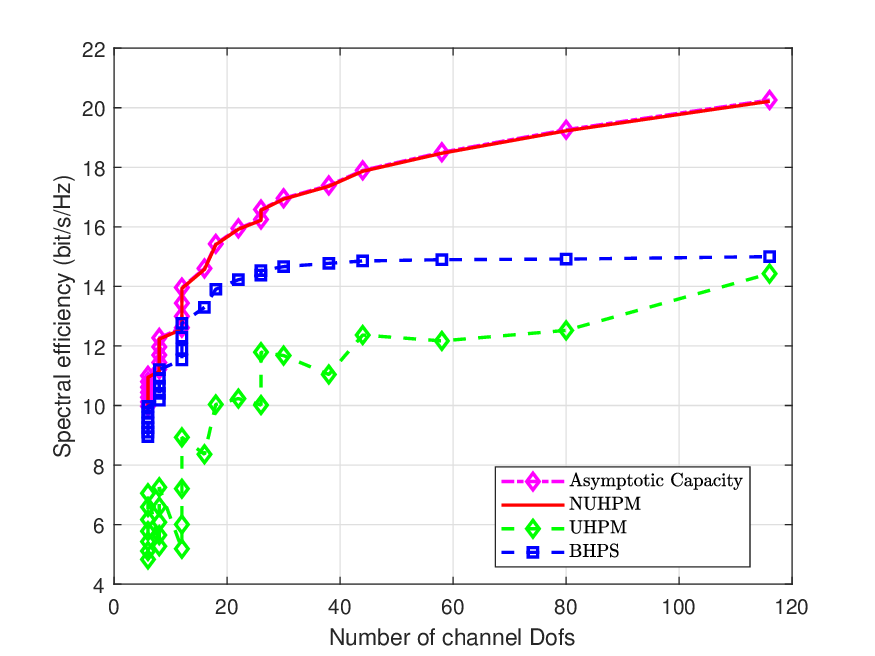}
    \caption{Spectral efficiency versus the number of DoFs of the electromagnetic-based channel at an SNR of $20$ dB.}
    \label{sedof}
\end{figure}
\subsection{Advantages in Employing Varying Number of DoFs}
As discussed in the introduction, the system performance is influenced by the DoF. However, different beamforming designs and system settings result in different relationships between the performance metrics and DoFs. In this subsection, we carry out numerical simulations to assess the spectral efficiency for both schemes in terms of the number of spatial DoFs. As shown in Fig. \ref{sedof}, NUHPM can utilize DoFs to approach the performance limits of holographic MIMO systems more effectively compared to BHPS {and UHPM}. Moreover, it is interesting to note that BHPS arrives at its maximum spectral efficiency with around 40 DoFs and then keeps nearly unchanged, but the performance under NUHPM continues to be improved as the number of DoFs increases, although with a gradually lower trend. In this case, the performance gap between NUHPM and BHPS grows larger and larger. The reason behind this phenomenon is NUHPM utilizes all available DoFs to deliver information, while BHPS only cares about the best propagation channels. It indicates that in a green communication system with limited RF chains, NUHPM is a much better choice, as the BHPS cannot fully exploit the advantages provided by additional spatial DoFs.

\subsection{Performance Improvement by Only Increasing the Transmitting Antenna Area}

\begin{figure}
    \centering
    \includegraphics[width=1\linewidth]{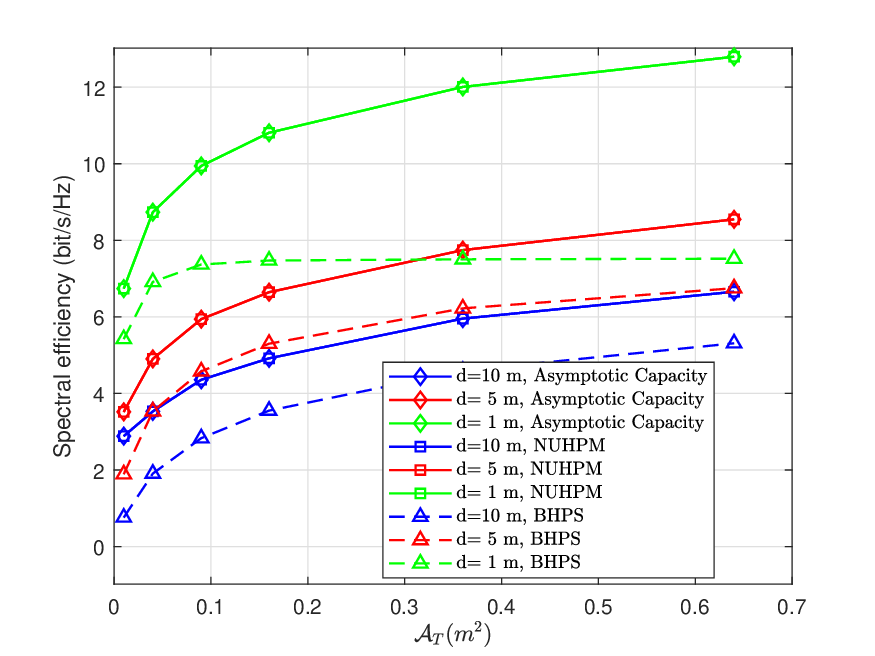}
    \caption{Spectral efficiency versus the size of transmitting antenna surface at an SNR of $20$ dB.}
    \label{size}
\end{figure}
\begin{figure}
    \centering
    \includegraphics[width=1\linewidth]{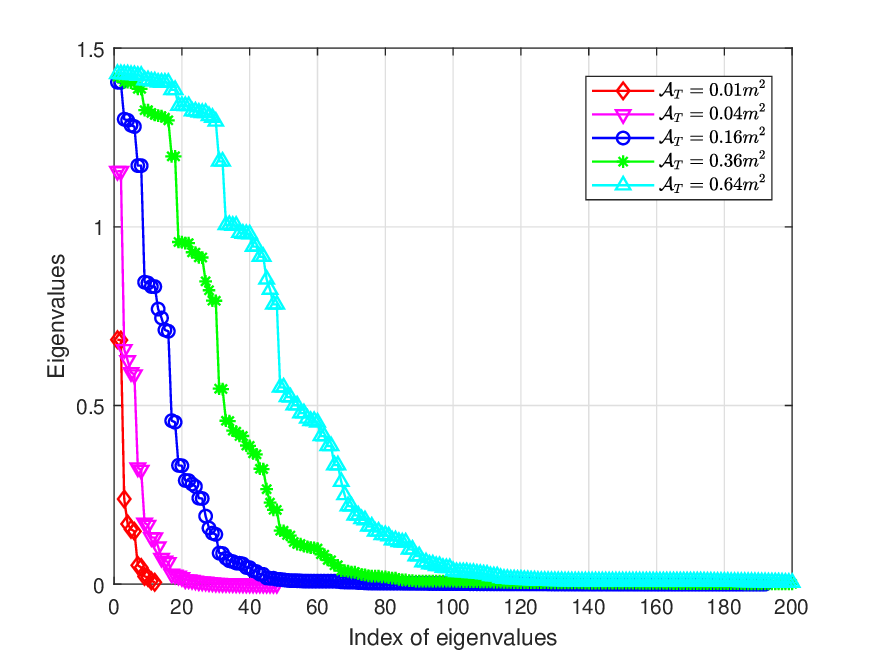}
    \caption{Distribution of channel eigenvalues for different antenna sizes, $d$=1 m.}
    \label{eig}
\end{figure}
To demonstrate the performance of the green evaluation by only increasing the size of the antenna aperture without increasing the DoFs, Fig. \ref{size} plots the spectral efficiency as a function of $\mathcal{A}_t$ when $d$ = 1, 5 and 10 m, in which the size of the receiver surface remains the same as the above simulation\footnote{Since we assume that the size of the patch antenna is fixed, the area $\mathcal{A}_t$ can also be understood as the number of transmitted antennas.}. In addition to already known finds that NUHPM is superior to BHPS, increasing the antenna area always brings performance gain for NUHPM, but not the same case for BHPS. It is worth noting that in the communication scenarios with short propagation distance, taking $d=1$ m for granted, increasing the antenna area can hardly bring gains to BHPS. Moreover, the spectral efficiency of both schemes saturate more sharply as $\mathcal{A}_T$ 
 continually shrinks. These findings leave us with an important insight that NUHPM can increase the capacity by increasing the antenna aperture (area) without increasing the number of RF chains. 

In order to double-check the positive role of antenna aperture on the NUHPM-supported holographic MIMO systems, we investigate the distribution of eigenvalues in Fig. \ref{eig}. According to (\ref{CH}) and (\ref{CB}), it can be found that when $K = 1$, the capacity of BHPS is directly determined by the largest eigenvalue, while the capacity of NUHPM is positively scaled with the sum of all eigenvalues. We know that the channel DoF is defined as the number of eigenvalues far away from $0$. It can be seen from Fig. \ref{eig} that although the DoF increases with the expansion of the antenna area, the maximum eigenvalue hardly changes, so the capacity of BHPS quickly reaches saturation, which can also explain why the curve of BHPS in Fig. \ref{sedof} does not change at high DoFs. Furthermore, Fig. \ref{eig} also shows that as the antenna area increases, the number of larger eigenvalues also increases, so NUHPM can increase the capacity by increasing the antenna area, especially when the distance between the transceiver is close.

\subsection{Impact of The Number of RF Chains at Transceivers}

\begin{figure}
    \centering
    \includegraphics[width=1\linewidth]{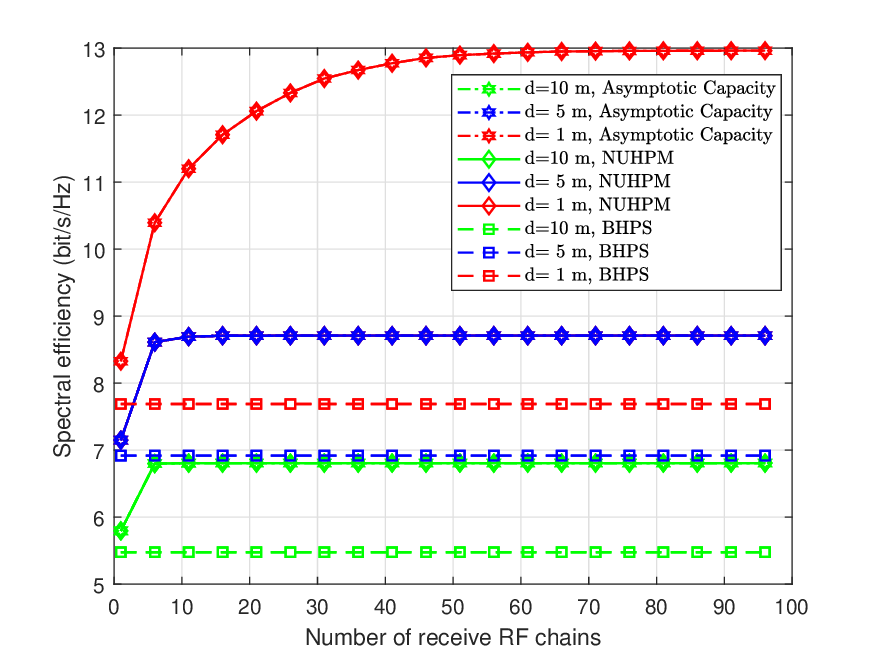}
    \caption{Spectral efficiency versus the number of RF chains at receiver with an SNR of $20$ dB.}
    \label{Nrrf}
\end{figure}
\begin{figure}
    \centering
    \includegraphics[width=1\linewidth]{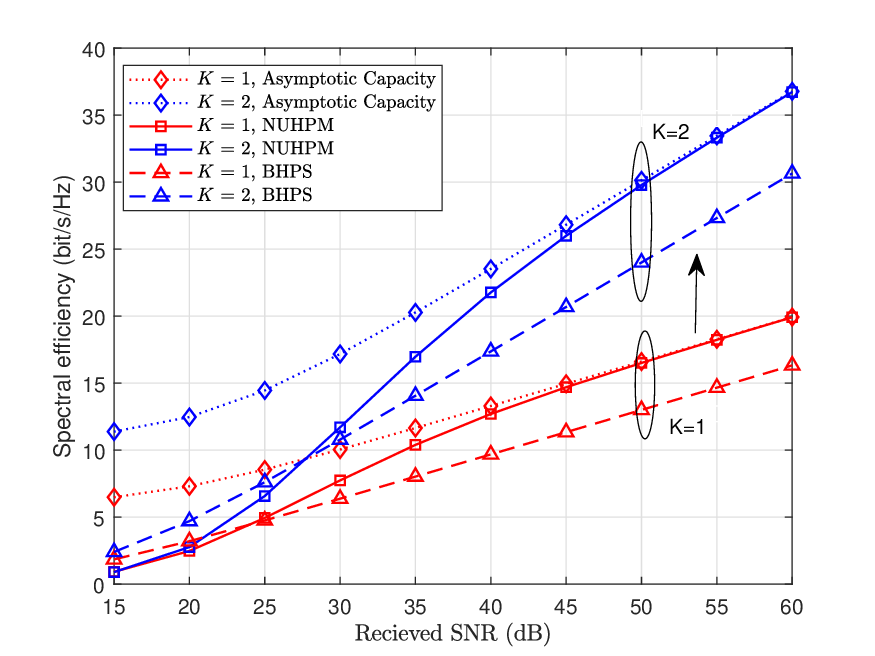}
    \caption{Spectral efficiency versus the  received SNR with different numbers of transmit RF chains.}
    \label{Ntrf}
\end{figure}
In this subsection, we present the impact of different numbers of transceiver RF chains on spectral efficiency. Firstly, we fix the number of transmitted RF chains $K=1$, change the number of RF chains at the receiver, and perform comparisons over different transmission distances, that is, $d = 1,5,10$ m. The simulation result is shown in Fig. \ref{Nrrf}. Since BHPS selects the optimal sub-channel for transmission, as long as the number of received RF chains end is not less than $K$, its variation does not affect the capacity a lot. However, this is not the same case for NUHPM, as it selects all sub-channels with non-zero eigenvalues for transmission. In this case, the number of received RF chains means the number of sub-channels that can be selected for reception. It can observe from Fig. \ref{Nrrf} that the closer the distance, the higher the DoF, the higher the number of subchannels, and the greater the impact of the number of received RF chains on the capacity under NUHPM. Therefore, a reasonable selection of the number of received RF chains can effectively improve the capacity while remaining energy-friendly. Most studies assume that the receiving RF chains are sufficient, that is, equal to the number of receiving antennas, which is not only energy-cost but also does not bring corresponding performance gains. These findings can not only help us to utilize NUHPM to enhance the system performance but also cut unnecessary receiving RF chains, thereby saving hardware resources and energy consumption. On the other hand, the number of working RF chains can be adaptively adjusted according to the performance requirements under the resources or energy constraint. These valuable insights facilitate the realization of GH-MIMO communications in practice. 

In Fig. \ref{Ntrf}, we then show the impact of the number of transmitting RF chains when assuming the communication distance between the transmitting surface and receiving surface is $d = 2$ m, and the RF chains at the receiver are sufficient. It can be seen that the performance gap between NUHPM to BHPS grows from $4$ bit/s/Hz to $7$ bit/s/Hz when there are additional transmitting RF chains considering around $50$ dB received SNR. It indicates that the increment in RF chains indeed brings performance gain for holographic MIMO communications, but the NUHPM-supported system is worth the increase over the BHPS-supported system. 

In summary, the above-discussed system parameters' variations, such as increasing DoF, enlarging antenna aperture, and reducing propagation distance, can all enhance the system performance under both schemes. However, NUHPM always enjoys more performance improvement than BHPS, which verifies the effectiveness of the proposed scheme. The achieved fancy performance also proves the possibility to realize the required Holographic MIMO system in a green way with only a few RF chains.

\subsection{Evaluating the Performance of Energy Efficiency}
\begin{figure}
    \centering
    \includegraphics[width=1\linewidth]{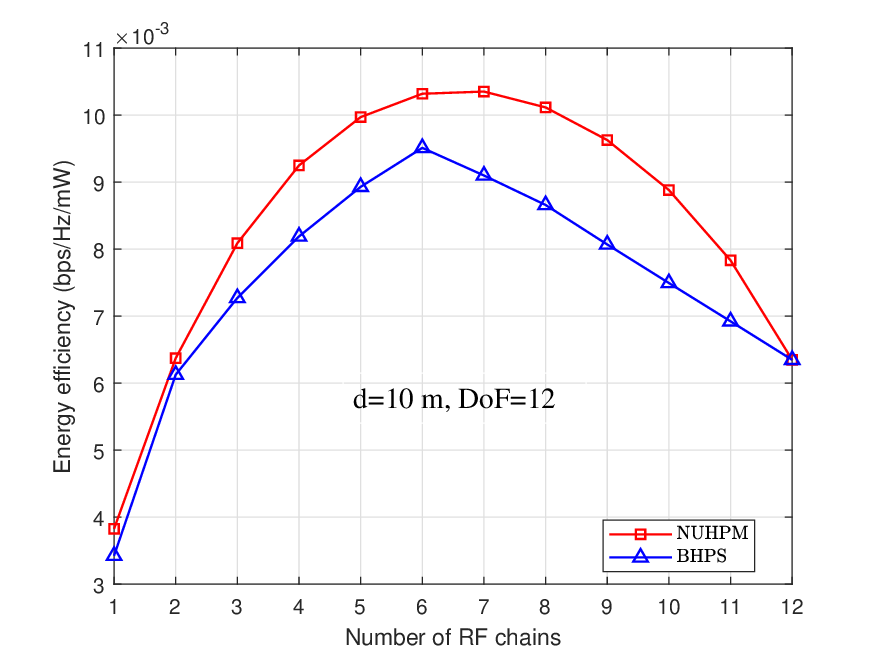}
    \caption{Energy efficiency versus numbers of transmit RF chains, $d=10$ m.}
    \label{Nrf}
\end{figure}

The above simulations demonstrate that our proposed NUHPM achieves higher spectral efficiency compared to BHPS. This is attributed to the flexible pattern activation, which effectively utilizes the channel DoFs and improves the utilization of physical resources. However, it is also crucial to consider energy efficiency as another key performance factor in the simulation to highlight the green performance of NUHPM. 
For simplicity, we claim a downlink power consumption model at the base station, and the energy efficiency can be formulated as
\begin{equation}
    \label{}
    \eta=\frac{C}{P_T+KP_{RF}},
\end{equation}
where $P_{T}$ is the common power of the transmitter, $P_{RF}$ is the power of each RF chain. We assume that the total power consumption $P_{total}=P_T+KP_{RF}=3000$ mW and $P_{RF}=160$ mW. In Fig. \ref{Nrf}, we plot the curves showing the energy efficiency of NUHPM and BHPS as the number of RF chains varies when the distance is $10$ m. It can be found that as the RF chain increases, the two curves first increase and then decrease. That is because the RF chain increase causes the transmitter power to decrease and the trade-off phenomenon occurs. Meanwhile, the energy efficiency of NUHPM peaks near $K=6$, when the size of the pattern candidate set $|\mathcal{E}|$ is the largest and the utilization of resources is the highest. When $K=DoF=12$, $|\mathcal{E}|=1$, then NUNPM is same as BHPS. However, what is certain is that NUHPM energy efficiency performance is better than BHPS and is more suitable for green communications.

\section{Conclusions}
In this paper,  we investigated GH-MIMO communications for green and sustainable MIMO air interface evaluation. In GH-MIMO, only a few transmit RF chain is equipped and additional spatial DoFs are adopted as modulation resources for capacity improvements. We derived the closed-form asymptotic capacity. The optimal holographic pattern activation probability distribution and optimal data stream input distribution were derived.
Comprehensive simulations are conducted. Both theoretical and simulation results show that the proposed NUHPM scheme considerably outperforms the existing BHPS scheme. The NUHPM can flexibly employ a varying number of DoFs when the user moves. The impact of the numbers of transmit and receive RF chains on the capacity is also investigated.  Besides, we also demonstrated that the capacity can be considerably improved by only increasing the antenna area without increasing the number of RF chains. This can facilitate a lot of applications in future green communication networks.

\appendices
\section{Proof of Proposition 1}
\begin{IEEEproof}
Substituting (\ref{eq22}) into (\ref{eq16}) yields
\begin{equation}\label{eq30}
\lim_{1/\sigma^2\rightarrow +\infty} \mathsf{I}(\hat{\mathbf{x}};\hat{\mathbf{y}}|\mathbf{\Lambda})=\lim_{1/\sigma^2\rightarrow +\infty} {\mathsf{H}_{LB}(\hat{\mathbf{y}})}-N-N\log_2 \pi.
\end{equation}
Recalling the expression of $\mathsf{H}_{LB}(\hat{\mathbf{y}})$ in (\ref{eq21}) and substituting it in (\ref{eq30}), we obtain
\begin{equation}
\begin{split}
&\lim_{1/\sigma^2\rightarrow +\infty} \mathsf{I}(\hat{\mathbf{x}};\hat{\mathbf{y}}|\mathbf{\Lambda})\\&= -\sum_{i=1}^{|\mathcal{E}|}p_i\log_2\left(\sum_{j=1}^{|\mathcal{E}| }\frac{p_j}{\pi^{N}\det(\mathbf{D}_i+\mathbf{D}_j)}\right)-N-N\log_2 \pi.
\end{split}
\end{equation}
By extracting  $\log_2(1/\pi^N)$ out of the first term, we derive
\begin{equation}\label{eq32}
\begin{split}
&\lim_{1/\sigma^2\rightarrow +\infty} \mathsf{I}(\hat{\mathbf{x}};\hat{\mathbf{y}}|\mathbf{\Lambda})\\&=-\sum_{i=1}^{|\mathcal{E}|}p_i\log_2\left(\sum_{j=1}^{|\mathcal{E}| }\frac{p_j}{\det(\mathbf{D}_i+\mathbf{D}_j)}\right)-N
\\&=-\sum_{i=1}^{|\mathcal{E}|}p_i\log_2\left(\frac{p_i}{\det(2\mathbf{D}_i)}+\sum_{j\neq i}^{|\mathcal{E}| }\frac{p_j}{\det(\mathbf{D}_i+\mathbf{D}_j^*)}\right)-N.
\end{split}
\end{equation}
Due to the fact $\det(\mathbf{D}_i+\mathbf{D}_j)\propto(1/\sigma^2)^{N-K}$ (c.f., [36, Eq. 41]) and $N>K$, thus
\begin{equation}\label{eq50}
\lim_{1/\sigma^2\rightarrow +\infty}\sum_{j\neq i}^{|\mathcal{E}| }\frac{p_j}{\det(\mathbf{D}_i+\mathbf{D}_j)}=0.
\end{equation}
Substituting (\ref{eq50}) into (\ref{eq32}) yields
\begin{equation}\label{eq24b}
\begin{split}
&\lim_{1/\sigma^2\rightarrow +\infty} \mathsf{I}(\hat{\mathbf{x}};\hat{\mathbf{y}}|\mathbf{\Lambda})\\
&=-\sum_{i=1}^{|\mathcal{E}|}p_i\log_2\left(\frac{p_i}{\det(2\mathbf{D}^*_i)}\right)-N\\
&=-\sum_{i=1}^{|\mathcal{E}|}p_i\log_2 {p_i}+\sum_{i=1}^{|\mathcal{E}|}p_i\log_2\det(2\mathbf{D}_i)-N\\
&=-\sum_{i=1}^{|\mathcal{E}|}p_i\log_2{p_i}+\sum_{i=1}^{|\mathcal{E}|}p_i\log_2\left[2^{N}\det(\mathbf{D}_i)\right]-N\\
&=\sum_{i=1}^{|\mathcal{E}|}p_i\log_2\det\left(\mathbf{D}_i\right)-\sum_{i=1}^{|\mathcal{E}|}p_i\log_2p_i.
\end{split}
\end{equation}
\end{IEEEproof}
\section{Proof of Theorem 1}
\begin{IEEEproof}
Substituting (\ref{eq15}) into problem (\ref{eq24}) yields
\begin{equation}\label{OP}
\begin{split}\max_{\{p_i\},\{\mathbf{Q}_i\}}&\sum_{i=1}^{|\mathcal{E}|}p_i\log_2\det\left(\mathbf{I}_{N}+\frac{1}{\sigma^2}\mathbf{\Lambda}\mathbf{E}_i\mathbf{Q}_i\mathbf{E}_i^H\mathbf{\Lambda}^H\right)\\&-\sum_{i=1}^{|\mathcal{E}|}p_i\log_2p_i\\
\mathrm{s.~t.:}& \sum_{i=1}^{|\mathcal{E}|}p_i\tr(\mathbf{Q}_i)=1,~\sum_{i=1}^{|\mathcal{E}|}p_i=1.
\end{split}
\end{equation}

Let $\mathbf{\Lambda}_{i}=\diag(\lambda_{i1},\cdots,\lambda_{iK})\in\mathbb{C}^{K\times K}$, where $\lambda_{i1},\cdots,\lambda_{iK}$ denote the $K$ singular values selected by $\mathbf{E}_i$. Then, the optimization problem in (\ref{OP}) can be written as
\begin{equation}\label{eq47}
\begin{split}
\max_{\{p_i\},{\{\mathbf{Q}_i\}}}&\sum_{i=1}^{|\mathcal{E}|}p_i\log_2\det\left(\mathbf{I}_{K}+\frac{1}{\sigma^2}\mathbf{\Lambda}_i{\mathbf{Q}_i}\mathbf{\Lambda}_i^H\right)\\&-\sum_{i=1}^{|\mathcal{E}|}p_i\log_2p_i\\
\mathrm{s.~t.:}& \sum_{i=1}^{|\mathcal{E}|}p_i\tr({\mathbf{Q}_i})=1,~\sum_{i=1}^{|\mathcal{E}|}p_i=1.
\end{split}
\end{equation}
We express 
$\mathbf{Q}_i$ by $\mathbf{Q}_{i}=\diag(\sigma_{i1},\cdots,\sigma_{iK})\in\mathbb{C}^{K\times K}$.  As the SNR goes high, the term $\log_2\det\left(\mathbf{I}_{N}+\frac{1}{\sigma^2}\mathbf{\Lambda}_i{\mathbf{Q}_i}\mathbf{\Lambda}_i^H\right)$ can be expressed as 
\begin{equation}
\begin{split}
&\lim_{{1}/{\sigma^2}\rightarrow\infty}\log_2\det\left(\mathbf{I}_{N}+\frac{1}{\sigma^2}\mathbf{\Lambda}_i{\mathbf{Q}_i}\mathbf{\Lambda}_i^H\right)\\&\approx\log_2\det\left(\frac{1}{\sigma^2}\mathbf{\Lambda}_i{\mathbf{Q}_i}\mathbf{\Lambda}_i^H\right)\\
&=\sum_{j=1}^{K}\log_2\frac{1}{\sigma^2}\sigma_{ij}\lambda_{ij}^2.
\end{split}
\end{equation}
With the total power of $K$ data streams $\hat{\mathbf{s}_i}$ being denoted by $b_i=\tr(\mathbf{Q}_i)$. Determining the optimal power $\{\sigma_{ij}\}$ for the $K$ data streams with budget $b_i$ is a typical power allocation problem for paralleled MIMO channels, and the optimal solution is the famous water-filling power allocation. In the high SNR regime, the optimal water-filling power allocation is approximately equal power allocation \cite{Goldsmith2003}, i.e., $\sigma_i=\frac{b_{i}}{K}$. Taking this into consideration, we get
\begin{equation}\label{NE59}
\begin{split}
&\lim_{{1}/{\sigma^2}\rightarrow\infty}\log_2\det\left(\mathbf{I}_{N}+\frac{1}{\sigma^2}\mathbf{\Lambda}_i{\mathbf{Q}_i}\mathbf{\Lambda}_i^H\right)\\&=\sum_{j=1}^{K}\log_2\frac{\frac{1}{\sigma^2} b_{i}\lambda_{ij}^2}{K}\\
&={K}\log_2 b_i+q_i,
\end{split}
\end{equation}
where $q_i\triangleq \sum_{j=1}^{{K}}\log_2\frac{\frac{1}{\sigma^2}\lambda_{ij}^2}{K}$ is a constant. By substituting (\ref{NE59}) into (\ref{eq47}), we can write the power allocation for different holographic activation as
\begin{equation}\label{eq49}
\begin{split}
\max_{\{b_i\}}&\sum_{i=1}^{|\mathcal{E}|}p_i\left[{K}\log_2 b_i+q_i-\log_2p_i\right]\\
\mathrm{s.~t.:}& \sum_{i=1}^{|\mathcal{E}|}p_ib_i=1,~\sum_{i=1}^{|\mathcal{E}|}p_i=1.
\end{split}
\end{equation}
 
To deal with the optimization problem,  we introduce the Lagrange function as
\begin{equation}\label{eqL3}
\begin{split}
&J(\{b_i\},\zeta)
=\\&\sum_{i=1}^{|\mathcal{E}|}p_i\left[ {K}\log_2 b_i+q_i+\log_2 p_i\right]-\zeta\left(\sum_{i=1}^{|\mathcal{F}|}p_ib_i-1\right).
\end{split}
\end{equation}
 By calculating the  partial  derivation of the Lagrange function regarding $b_i$,  we obtain
\begin{equation}
\frac{{K}p_i}{b_i}-\zeta p_i=0,~i=1,\cdots, |\mathcal{E}|.
\end{equation}
Then, we obtain
\begin{equation}
b_1=\cdots=b_{|\mathcal{E}|}=\frac{K}{\zeta}.
\end{equation}
Owing to $\sum_{i=1}^{|\mathcal{E}|}p_ib_i-1=0$,  we derive the optimal solution to be
\begin{equation}
b_1^*=\cdots=b_{|\mathcal{E}|}^*=1.
\end{equation}
Then, the diagonal elements $\{\sigma_{ij}^*\}$ of the optimal ${\mathbf{Q}_i^*}$ as the water-filling power allocation can be obtained as that in Theorem 1.

Given $\{\mathbf{Q}_i^*\}$ with $\tr\{\mathbf{Q}_i^*\}=1$, the Lagrange function of the problem in (\ref{OP}) can be formulated as
\begin{equation}\label{eqL1}
\begin{split}
\mathcal{L}(\{p_i\},\mu)=\sum_{i=1}^{|\mathcal{E}|}p_i\left( \log_2\det(\mathbf{D}_i^*)-\log_2p_i\right)-\mu\left(\sum_{i=1}^{|\mathcal{E}|}p_i-1\right).
\end{split}
\end{equation}
It is worthy mentioning that the power constraint $\sum_{i=1}^{|\mathcal{E}|}p_i\tr(\mathbf{Q}_i^*)=1$ in (\ref{OP}) is omitted as it always holds.

Taking the partial derivations of the Lagrange function in (\ref{eqL1}) with respect to $p_i$ yields
\begin{equation}
\log_2\det(\mathbf{D}_i^*)-\log_2 p_i-1/\ln2 -\mu=0,~i=1,\cdots,|\mathcal{E}|.
\end{equation}
By solving the above equations, we obtain
\begin{equation}
p_i=\frac{2^{-\mu}\det(\mathbf{D}_i^*)}{e},~i=1,\cdots,|\mathcal{E}|.
\end{equation}
Considering the constraint $\sum_{i=1}^{|\mathcal{E}|}p_i=1$, the optimal $p_i$ can be derived as
 \begin{equation}
p_i^{*}=\frac{\det(\mathbf{D}_i^*)}{\sum_{i=1}^{|\mathcal{E}|}\det(\mathbf{D}_i^*)},~i=1,\cdots,|\mathcal{E}|.
\end{equation}
Then, the proof of Theorem 1 is complete. 

\end{IEEEproof}

\bibliographystyle{IEEEtran} 
\bibliography{IEEEabrv,bib}

\end{document}